\documentclass{article}
\usepackage{amsmath,amssymb}
\usepackage{latexsym,graphicx,epsfig,wrapfig}
\usepackage{latexsym,graphicx,epsfig}
\usepackage{wrapfig}

\def\lb{\linebreak[4]}
\newcommand{\be}{\begin{equation}}
\newcommand{\ee}{\end{equation}}
\newcommand{\bes}{\begin{subequations}}
\newcommand{\ees}{\end{subequations}}
\newcommand{\bea}{\begin{eqnarray}}
\newcommand{\eea}{\end{eqnarray}}
\newcommand{\bear}{\begin{equation}\begin{array}}
\newcommand{\eear}[1]{\end{array}\label{#1}\end{equation}}
\def\ba{$$\begin{array}}
\def\ea{\end{array}$$}
\def\bra{$\begin{array}}
 \def\era{\end{array}$}

\newcommand{\fr}[2]{\dfrac{{ #1}}{{ #2}}}

\def\vep{{\varepsilon}}
\newcommand{\epe}{\mbox{$e^+e^-\,$}}
\newcommand{\ggam}{\mbox{$\gamma\gamma\,$}}

\def\cl{\centerline}

\newsavebox{\fmbox}

{\end{list}}
\newcounter{enumct}

\newcommand{\bu}{$\bullet$\ }

\begin{document}
\title{ About earlier history of two--photon physics}
\author{{\it I. F. Ginzburg,}\\
{ Sobolev Institute of Mathematics,  Novosibirsk, Russia}}

\date{}
\maketitle

\cl{\it To be published in Proc. Photon05, Acta Physica
Polonica}

\begin{abstract}
The earlier history of two--photon physics is reviewed.
\end{abstract}

\maketitle

The term  {\bf two--photon processes} is used now for the
reactions  in which some system of particles is produced in
collision of two photons, either real or virtual. In the
study of these processes the principal goal is to describe
main features of proper two--photon process separating it
from mechanism which responsible for the production of
incident photons.

\bu An early interest in the two--photon physics has arisen
after discovery of positron by Anderson (1932). There
appeared a necessity to find out the process in which
positrons are generated. In 1934 studying \epe\, pair
production in collision of ultrarelativistic charged
particles Landau and Lifshitz \cite{LL} have ascertained
that the two photon channel of Fig.~1 is dominated here.
\begin{wrapfigure}[10]{l}[0pt]{4.5cm}\vspace{-2mm}
\includegraphics[width=0.4\textwidth,
  height=6.5cm]{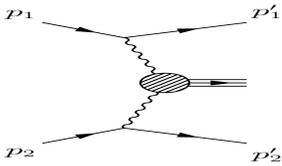}\\[-48mm]
    \caption{Two photon particle production in collision of two
fast particles.}
\end{wrapfigure}
They calculated the cross section of the process $Z_1Z_2\to
Z_1 Z_2 e^+e^-$ in the leading logarithmic approximation.
Almost simultaneously Bethe and Heitler \cite{BH}
considered \epe\, pair production by photon in the field of
a nuclei. This process contain subprocess $\ggam\to\epe$,
like Fig.~1.

The leading log result of \cite{LL} was improved by Racah
\cite{Racah} who have calculated the corresponding cross
section with a high accuracy   $\sim (M/E)^2$ where $E$ and
$M$ are energy and mass of incident nuclei. The process
$\gamma A\to \epe A$ was included in the theory of wide
atmospheric showers in cosmic rays \cite{LRum} and in the
description of the energy losses of fast muons in matter
\cite{enloss}.

The two-photon hadron production was considered for the
first time by Primakoff \cite{Prim} suggested in 1951 to
measure the $\pi^0$ life--time in the reaction $\gamma Z\to
\pi^0 Z$. The new interest to such processes appeared when
the construction of \epe\  colliders become close to a
reality. In 1960 Low \cite{Low} paid attention to the fact
that the $\pi^0$ life--time can be measured also in in
\epe\, collisions. Simultaneously the two--photon reaction
$\epe\to\epe\pi^+\pi^-$ (for point--like pions) was
considered \cite{CalZem}. However, the rates involved
seemed unmeasurable at that time and no further work was
done.

In 1969--1970 new generation of papers appeared with the
goal to cover possible set of final states of \epe\
colliders as complete as possible. Authors considered the
two--photon production of $\pi^0$, $\eta$ and point-like
pion and kaon pairs \cite{deCel}. Some of these processes
and purely QED processes $\epe\to\epe\pi^+\pi^-$,
$\epe\to\epe\mu^+\mu^-$ were considered in more detail by
Paris \cite{Kess} and Novosibirsk BINP \cite{Bai} groups.
These papers did not provoke high interest in particle
physics community since they were in line with numerous
calculations of various processes at \epe\  colliders with
small cross section (at the contemporary energies) and
don't pretend for obtaining of new information except new
tests of QED.

\bu To the end of 1969,  results in the study of deep
inelastic $ep$ scattering were a hot point in particle
physics. Besides, a preliminary information about
experimental discovery of $\epe\to\epe\epe$ process in
Novosibirsk BINP \cite{Bal} became known. Under influence
of these two facts Novosibirsk IM group wrote the paper
submitted to Russian Pis'ma ZhETF at May 4 and appeared
there at June 5, 1970 \cite{BBG}, Fig.2  (it was translated
soon).
\begin{figure}[htb]
\begin{center}
  \includegraphics[width=0.95\textwidth,
  height=6.5cm]{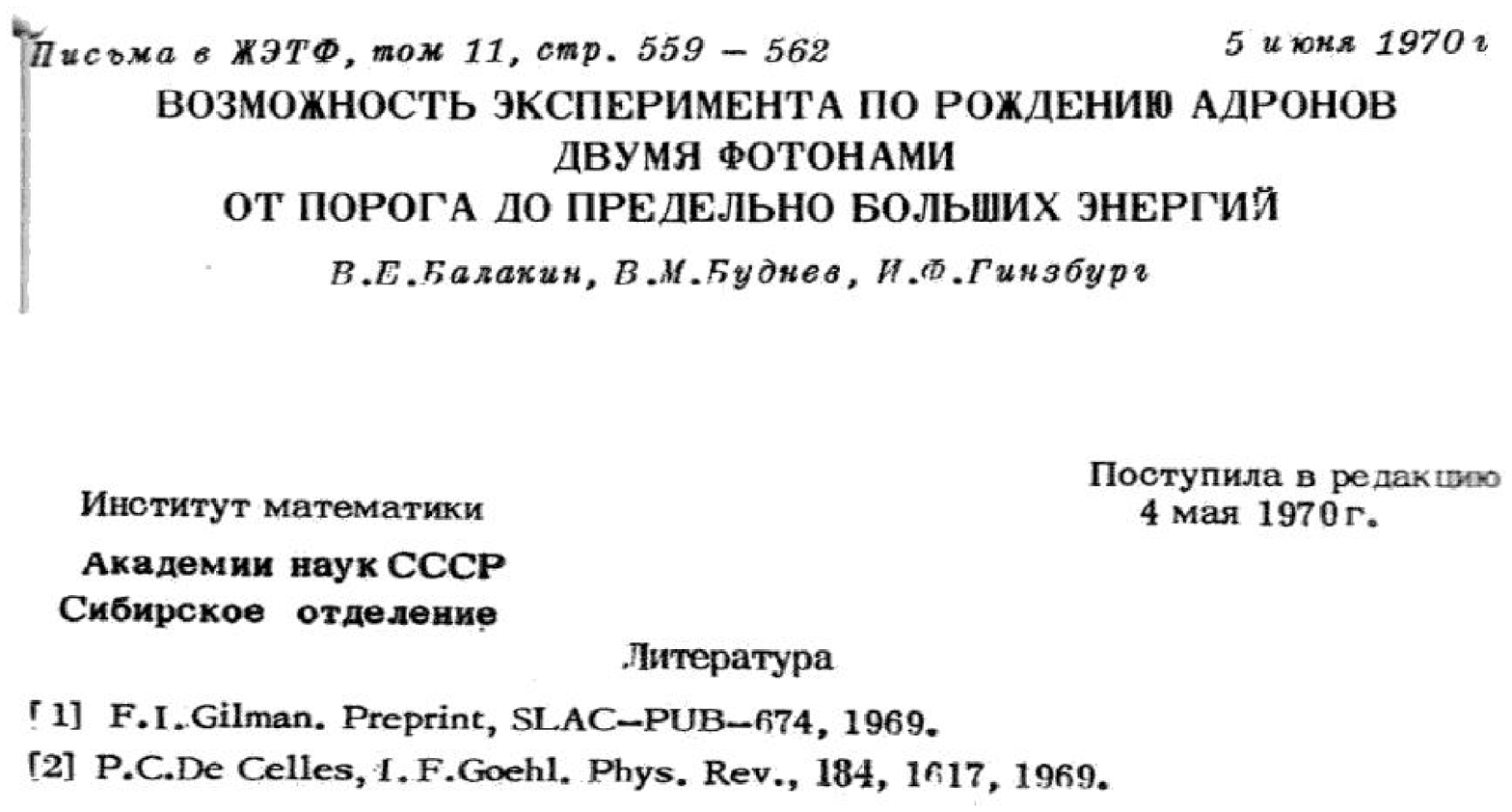}
    \caption{V.E. Balakin, V.M. Budnev, I.F. Ginzburg,
"{\it The possibility of experiment with production of
hadrons by two photons from threshold to the extremely high
energies}", published June 5, 1970, submitted May 4, 1970,
Russian Pis'ma ZhETF}
\end{center}
\end{figure}
Here it was shown that the experiments at \epe colliders
{\it open new experimental field of particle physics} --
the opportunity to  extract from the data an information
about fundamental process, $\gamma^*\gamma^*\to hadrons$
{\it (or some other particles)}. The paper contains also
estimate of high energy total cross section\lb
$\sigma(\ggam\to\,hadrons)\sim \sigma^2(\gamma
p)/\sigma(pp)\sim (0.3\div 1)\mu$b, which is in accord with
modern measurements, and the equations for extraction of
two-photon cross sections from the data at small electron
scattering angles in the form which is used for this aim up
to now,
 \bear{c}
\fr{d\sigma}{dE_1dE_2d\Omega_1d\Omega_2}=
\left(\fr{\alpha}{2\pi^2}\right)^2\fr{1}{q_1^2q_2^2}
\fr{E_1E_2}{E^2}\,\fr{(E^2+E_1^2)(E^2+E_2^2)}{(E-E_1)(E-E_2)}
\sigma^{\gamma\gamma}_{exp}\,,\\[2mm]
\sigma^{\gamma\gamma}_{exp}= \sigma^{\gamma\gamma}_{TT}+
\vep_1\sigma^{\gamma\gamma}_{ST}+\vep_2\sigma^{\gamma\gamma}_{TS}+
\vep_1\vep_2\left(\sigma^{\gamma\gamma}_{SS}+
\tau_{TT}^{ex}\cos2\phi/2\right)
+\vep_3\tau_{TS}^{ex}\,,\\[2mm]
\vep_1\!=\!\fr{2EE_1}{E^2\!+\!E_1^2},\;\;\vep_2\!=\!
\fr{2EE_2}{E^2\!+\!E_2^2},\;\; \vep_3\!=
\!\vep_1\vep_2\fr{(E\!+\!E_1)(E\!+\!E_2)}{32E\sqrt{E_1E_2}}
\cos\phi.
 \eear{}
($E$ and  $E_i$ are the energies of initial and scattered
electrons, $\phi$ is the angle between their scattering
planes, other notations was not practically changed during
34 years.) The numerical estimates of anticipated cross
sections were done and it was found that the cross section
grows fast with beam energy. Besides, the sketch of
experimental program was formulated. In July this paper was
reported at Kiev Rochester conference (preliminary version
of paper \cite{BKT} was also reported there) -- see {\it
abstracts of Kiev--Rochester, 1970}.

Three month later after \cite{BBG},
\begin{figure}[htb]
\begin{center}
  \includegraphics[width=0.95\textwidth,
  height=8.6cm]{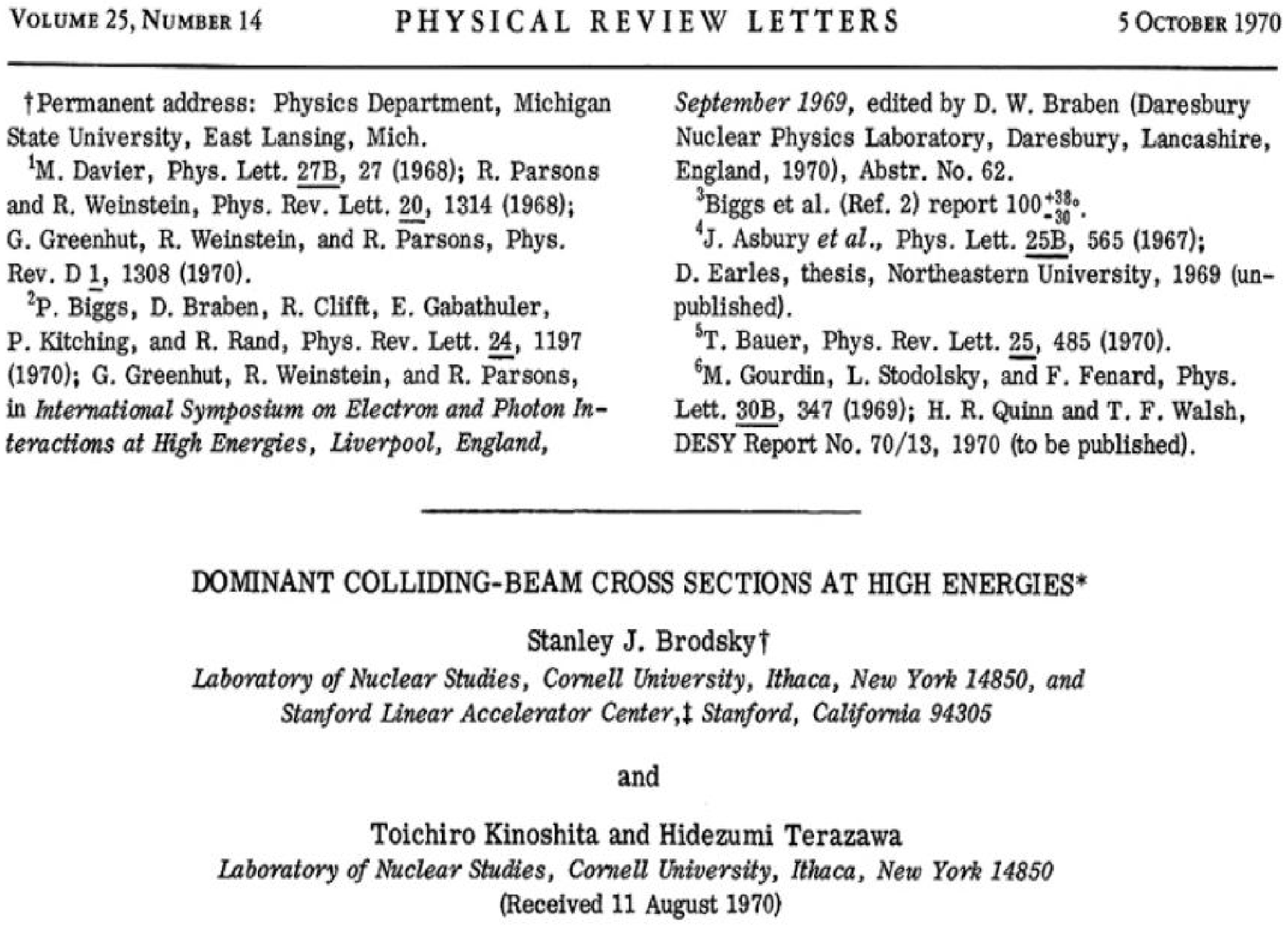}
\caption{S.J. Brodsky, T. Kinoshita, H. Terazawa, "{\it
Dominant colliding beam cross sections at high energies}",
published October 5, 1970, submitted August 11, 1970,
Physical Review Letters}
\end{center}
\end{figure}
S. Brodsky, T. Kinoshita, S. Terazawa have submitted in
Physical Review Letters their paper \cite{BKT} (Fig.3).
They consider two--photon production of $\pi^0$, $\eta$,
point-like $\pi^+\pi^-$ in \epe and $e^-e^-$ colliding
beams. They found that these cross sections grow fast with
beam energy and described some features of the angular
distributions of produced pions (in point-like QED
approximation). Analogously to \cite{BBG}, these results
had shown  that two-photon physics provides a large field
for theoretical studies and experimentation. They obtain
results with the aid of Weizsacker--Williams method with
incorrect spectra of equivalent photons (about twice larger
than correct for each).  Many authors of subsequent papers
reproduced this inaccuracy.

In 1971 VEPP-2 (BINP, Novosibirsk \cite{Bal}), and in 1972
ADONE (Frascati, Italy) \cite{ADONE} reported about the
observation of $\epe\to \epe\epe$ process.

\bu The papers \cite{BBG}--\cite{Bal} open window for
stream of publications devoted two--photon physics. The
theoretical publications considered different problems
related to details of data extraction, backgrounds, QED
processes and problems of hadron physics in \ggam\
collisions. The first stage of these studies was summarized
in review \cite{BGMS} containing all necessary equations
for data preparation and set of equations useful for
different estimates. This review contains also detail
description of equivalent photon (Weizsacker--Williams)
method, including estimate of its accuracy in different
situations. In 1974 authors of  \cite{GKST} cannot imagine
opportunity of longitudinal electron polarization at \epe\
storage rings and don't describe this case in basic
equations. This lacuna in \cite{BGMS} was closed in
\cite{GS}.

Most  of (theoretical) papers of 70-th devoted to hadron
physics in \ggam\ \  collisions were reproductions of
results and ideas considered earlier for other hadronic
systems. However it was found by Witten that the structure
function of photon is unique quantity in particle physics
which can be found from QCD at large enough $Q^2$ and $s$
completely without phenomenological parameters
\cite{Witten}. The verification of this result in future
experiments is necessary to verify that the QCD is indeed a
theory of strong interactions.

The real experimental activity in this field started, in
fact, in 1979 by SLAC experiment  in which it was
demonstrated that two--photon processes can be successfully
studied at the modern detectors without recording of the
scattered electron and positrons -- via the separation of
events with the small total transverse momentum of produced
system \cite{ATel79}. Beginning from this work, the
investigation of two--photon processes become essential
component of physical studies at each \epe collider. A
number of results obtained  are summarized for example in
Particle Data Review \cite{PDG}.

\bu Very new opportunities for two-photon physics were
found in 1981. It was shown that the creation of very high
energy linear $e^{\pm}e$ colliders will allow to transform
them into the $\gamma \gamma $ and $\gamma e$ colliders
with luminosity and energy close to those for the basic
$e^{\pm}e$ colliders (the  Photon Collider), with
relatively small additional equipment  \cite{GKST}.

In contrast with the photon collisions at \epe\ colliders,
having relatively small effective \ggam\ luminosity, Photon
Colliders will be competitive with other machines in the
discovery of New Physics effects. But that is quite another
history.\\

{\bf Acknowledgments.}  This research has been supported by
Russian grants RFBR 05-02-16211, NSh-2339.2003.2.

\end{document}